\documentclass[a4paper]{article}

\usepackage{graphicx}
\usepackage{amsmath}
\usepackage{dcolumn}%
\usepackage[colorlinks=true,linkcolor=blue,citecolor=blue,urlcolor=blue]{hyperref}
\usepackage{xr-hyper}
\usepackage[dvipsnames]{xcolor}
\usepackage{url}
\usepackage{upgreek}
\usepackage[LGR,T1]{fontenc}
\newcommand{\angstrom}{\textup{\AA}}
\usepackage[section]{placeins}
\usepackage{caption}
\usepackage{color,soul}
\usepackage{multirow}%
\usepackage{amsmath,amssymb,amsfonts}%
\usepackage{amsthm}%
\usepackage{mathrsfs}%
\usepackage[title]{appendix}%
\usepackage{xcolor}%
\usepackage{textcomp}%
\usepackage{manyfoot}%
\usepackage{booktabs}%
\usepackage{algorithm}%
\usepackage{algorithmicx}%
\usepackage{algpseudocode}%
\usepackage{listings}%
\usepackage{tabularx}
\usepackage[version=3]{mhchem} % Formula subscripts using \ce{}
\usepackage{siunitx}
\usepackage{mhchem}
\usepackage{setspace}
\usepackage{geometry}

\usepackage{authblk}
\usepackage{geometry}

\usepackage[scaled=0.95]{helvet}

\DeclareSIUnit\angstrom{\text {Å}}

\title{\textbf{General-purpose LLMs as Constrained Crystal Composition Generators}}

\author[1,*,$\dagger$]{Hedda Oschinski}
\author[1,*]{Maximilian L. Ach}
\author[1]{Konstantin S. Jakob}
\author[1,$\dagger$]{Christian Carbogno}
\author[1]{Karsten Reuter}
\date{}
\affil[1]{\textit{Fritz-Haber-Institut der Max-Planck-Gesellschaft, Faradayweg 4-6, D-14195 Berlin, Germany}}

\affil[*]{These authors contributed equally.}

\affil[$\dagger$]{E-mail: oschinski@fhi.mpg.de, carbogno@fhi.mpg.de}

\usepackage[url=false, isbn=false, eprint=false, articletitle=true, backend=biber,style=nature]{biblatex}
\begin{document}

\maketitle

\begin{abstract}

The targeted discovery of inorganic materials remains challenging due to the vastness of compositional design spaces and the high cost of exhaustive screening. Task-specific generative artificial intelligence represents a particularly efficient alternative to screening, yet demands tedious collection of training data before providing real benefit. General-purpose large language models (LLMs) have recently shown tremendous potential for the targeted generation of single, optimal materials compositions without the need for task-specific fine-tuning. 
However, it is unclear whether LLMs generally pose an advantage compared to specialized generative models, in particular in large design spaces.
Here, we demonstrate that such models are capable of covering entire regions of the targeted property space effectively and systematically. Using Elpasolite materials as an established benchmark for generative tasks in large chemical spaces, we find that an iterative prompt-and-response framework is able to recover on average 96\% of all low-energy Elpasolites in the target region. This performance, driven mainly by iterative in-context learning, surpasses the generative abilities of previous, task-specific models. Our results establish general-purpose LLMs as flexible and accessible components for inverse materials design workflows.

\end{abstract}

%%%MAIN TEXT%%%%

\section{Introduction}

The discovery of novel materials with desirable properties remains a central challenge in materials science and chemistry.\autocite{butler2018machine, sanchez2018inverse, park2024has, kneiding2026inverse}
Traditionally, this process relies on a combination of chemical intuition, expert knowledge, and laborious trial-and-error experimentation, an approach that is both time-consuming and somewhat reliant on serendipity.\autocite{hautier2012computer,liu2017materials}
The advent of computational methods, in particular density-functional theory (DFT), has enabled the systematic screening of candidate materials by predicting their properties quantitatively from first principles.\autocite{wang2021high,borlido2022computational}
Yet, even with modern high-throughput workflows the systematic exploration of combinatorially large composition spaces remains prohibitively expensive.
To overcome this bottleneck, data-driven approaches have emerged as powerful tools augmenting physics-based methods.\autocite{butler2018machine, cheng2025ai}
Machine learning models trained on existing materials databases can predict properties at a fraction of the computational cost,\autocite{xie2018crystal,batatia2025foundation} and generative models have been proposed to navigate vast chemical spaces more efficiently than by 
traditional sampling algorithms.
Among the approaches explored, generative adversarial networks (GANs), variational autoencoders (VAEs), reinforcement learning (RL), diffusion models, and autoregressive models have all shown promise in proposing novel compositions with specific target properties.\autocite{turk2022assessing, anstine2023generative, park2024has, du2024machine, gomez2018automatic, chen2025crystal, zeni2025generative, kneiding2026inverse} 
However, these models share common practical challenges: they require task-specific training on carefully curated datasets, demanding both substantial computational resources and domain expertise in model design and training.
Hence, their applicability is generally restricted to the particular material class and property on which they were trained.
Moreover, reliably enforcing physical and chemical constraints, such as charge neutrality, valence rules, or structural stability, is not straightforwardly achieved within these generative frameworks, possibly resulting in a non-trivial fraction of invalid or unphysical proposals.\autocite{jakob2026learning}

General-purpose large language models (LLMs) offer a different perspective.
Pre-trained on enormous and diverse corpora of text spanning the internet, books, and code -- of which scientific literature constitutes only a small fraction -- general-purpose LLMs have acquired a surprisingly broad chemical knowledge base, even without materials-specific fine-tuning.\autocite{van2025survey, zhang2025large}
A key advantage of such LLMs is their flexibility in representing materials (e.g., via plain-text descriptions), enabling straightforward application across various tasks without the need for fine-tuning.
In recent years, the materials science community has begun to make use of this for a range of problems, including property prediction, synthesis planning, and literature mining.\autocite{jablonka2024leveraging, guo2023can, tang2025matterchat, m2024augmenting, ock2023catalyst, lv2025bridging}
More broadly, tool-augmented and fully autonomous agentic systems that iterate through hypothesis generation, experimentation, and data analysis have been demonstrated across several scientific domains.\autocite{m2024augmenting, mitchener2025kosmos, boiko2023autonomous, nduma2025crystalyse, ghareeb2026multi, aygun2026ai, gottweis2026accelerating}

Beyond predictive applications, LLMs have also been applied to generative tasks: fine-tuned models have shown to produce stable inorganic crystal structures encoded as text,\autocite{gruver2024fine, flam2023language, antunes2024crystal} while pre-trained LLMs combined with evolutionary search strategies have been used to navigate chemical spaces for both organic molecules and crystalline materials.\autocite{wang2024efficient, sun2025synllama, gan2025matllmsearch}
LLMs have also been used to suggest and iteratively refine individual candidate compositions with respect to some target property, and early results are encouraging.\autocite{jia2024llmatdesign, sprueill2024chemreasoner, takahara2025accelerated}
These studies already support the idea that LLMs can be highly useful, particularly when the goal is to rapidly identify a small number of promising material candidates without the effort of generating large datasets and training dedicated models.

At the same time, it remains unclear if LLMs can systematically generate large numbers of chemically valid compositions that collectively cover a desired region of a property space.
For such applications, a large fraction of the community seems to intuitively assume that LLMs are inferior to problem-specific generative models. 
Compared to the generation of single, optimal materials, however, investigating the validity of this assumption and in turn the coverage of a certain design space is much more involved, as it in principle requires the properties of all materials in the design space to be known a priori.
As a result, exhaustive benchmarks on the generative capabilities of general-purpose LLMs for large design spaces are still lacking.  

\begin{figure*}[h!]
    \centering
    \includegraphics[width=\linewidth]{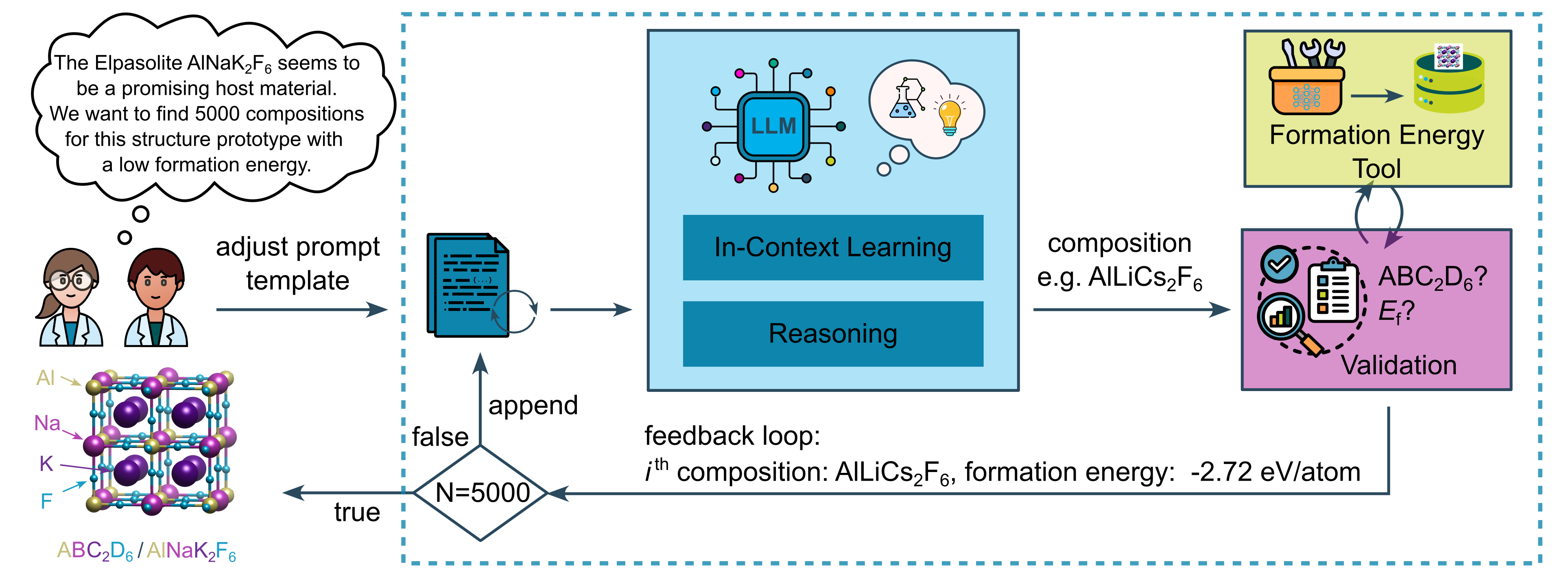}
    \caption{LLM-based generation workflow of low-formation-energy Elpasolite compositions. The LLM is prompted with generation instructions and uses its chemical intuition and reasoning to generate Elpasolite compositions. Compositions are validated for correct format, and a formation energy tool (in this case a database of pre-computed formation energies) determines the stability of the material. Finally, feedback is provided to the LLM depending on the formation energy, and the next composition is generated until the desired amount is met.}
    \label{fig:framework}
\end{figure*}

In this work, we address this issue using Elpasolite materials as a well-defined and pre-tabulated test system\autocite{faber2016machine, turk2022assessing} and demonstrate that general-purpose LLMs are indeed capable of systematically covering large composition spaces in a targeted fashion. 
Employing a conversational framework, as illustrated in Figure~\ref{fig:framework}, a general-purpose LLM (in this case GPT-5.4) iteratively proposes Elpasolite compositions and targets a formation energy below \SI{-2.26}{\electronvolt}/atom, a threshold met by under 0.2\% of the composition space.
We systematically investigate the factors that influence performance and find that general-purpose LLMs are comparable, if not superior to previously reported task-specific generative models.
Our analysis reveals that this is largely driven by the feedback-loop and not solely by the model's pre-existing chemistry knowledge.

\section{Results}

\subsection{Dataset and generation approach}
Elpasolites are quaternary double-perovskites with the general formula \ce{ABC2D6}. 
Their broad compositional flexibility has made them an important class of materials in functional materials research.
Halide perovskite materials such as \ce{Cs2AgBiX6} have for instance been thoroughly investigated for photovoltaic applications.\autocite{savory2016can,landini2022machine} 
In this context, the search for non-toxic, heavy metal-free, yet efficient materials requires exploration of large compositional spaces, as many different design aspects must be optimized.
In turn, stability descriptors such as the formation energy can serve as an initial filter to reduce the number of candidate materials.

Faber et al.\autocite{faber2016machine} conducted a large-scale screening of main-group Elpasolite formation energies across the compositional space of this crystal prototype. 
In their work, 39 main-group elements (H--Bi) were systematically substituted into the four crystallographic sites of the Elpasolite mineral, yielding close to 2~million possible compositions.
The formation energies of these compounds were predicted using kernel ridge regression trained on approximately 10{,}000 density functional theory calculations.
As a result, the dataset provides a comprehensive and well-characterized benchmark for computational materials discovery, particularly suited for evaluating generative models.

\begin{figure}[h!]
    \centering
    \includegraphics[width=0.6\linewidth]{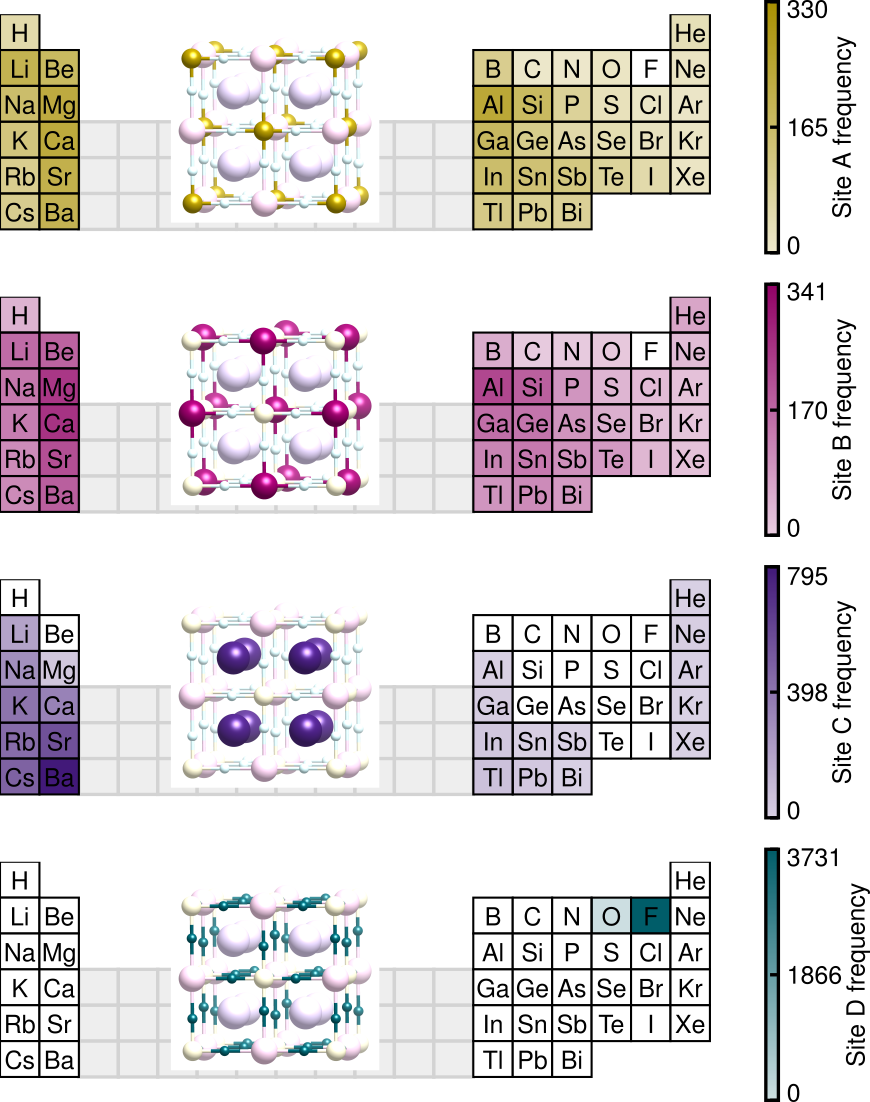}
    \caption{Illustration of the available crystallographic sites in the Elpasolite crystal structure, as well as the element distribution on each site for main-group Elpasolites with formation energies below $\SI{-2.26}{\electronvolt}/\text{atom}$, as identified by Faber et al.\autocite{faber2016machine}}
    \label{fig:pse_1}
\end{figure}

Since the complete chemical space is known, model performance can be assessed not only in terms of success rate, but also in terms of 
diversity, repetitiveness, and (sampling) efficiency.
Türk et al.\autocite{turk2022assessing} previously compared three deep generative models for this system --- 
a GAN, a VAE, and an RL model.
Their objective was to identify the 3740 Elpasolite compositions belonging to the lowest formation-energy class, characterized by $E_\mathrm{f} < \SI{-2.26}{\electronvolt}/\text{atom}$.
Within 5000 generation attempts, their models recovered 40--46\% of the target set, increasing to 75--94\% within 250{,}000 attempts.
Figure~{\ref{fig:pse_1}} depicts the four crystallographic sites in the Elpasolite unit cell, as well as the respective element distributions per site which yield formation energies below {$\SI{-2.26}{\electronvolt}/\text{atom}$}.

Building on this study, we employ the same dataset to assess the chemical knowledge and generative capabilities of general-purpose LLMs in direct comparison to these specialized architectures.
Rather than relying on task-specific training or fine-tuning, we adopt an iterative prompt-and-response workflow, in which the LLM acts as a composition generator within a broader discovery framework (see Figure~\ref{fig:framework}).
At each iteration, we instruct the model to propose a composition conforming to a predefined stoichiometric prototype \ce{ABC2D6} and target a desired property --- in this case, Elpasolite compositions with low formation energies.
The proposed candidate is then validated for compositional consistency, and the formation energy is retrieved either from a formation energy calculator or, as in the present study, a precomputed dataset.

Crucially, each generated composition and its associated formation energy are fed back to the model, creating a continuously expanding history of proposals.
This iterative feedback loop enables in-context learning across successive iterations, allowing the model to refine its search strategy without explicit parameter updates.
The resulting framework is simple, intuitive, and readily adaptable to different discovery tasks, while avoiding the overhead associated with training specialized generative models.
Full details of the prompting strategy and all hyperparameters are provided in the Supporting Information (Section~S1).
If not stated differently, we performed 3 runs for each setting and then display the average and the range of the best and the worst run.

\subsection{Performance comparison}

\begin{figure}[h]
    \centering
    \includegraphics[width=0.6\linewidth]{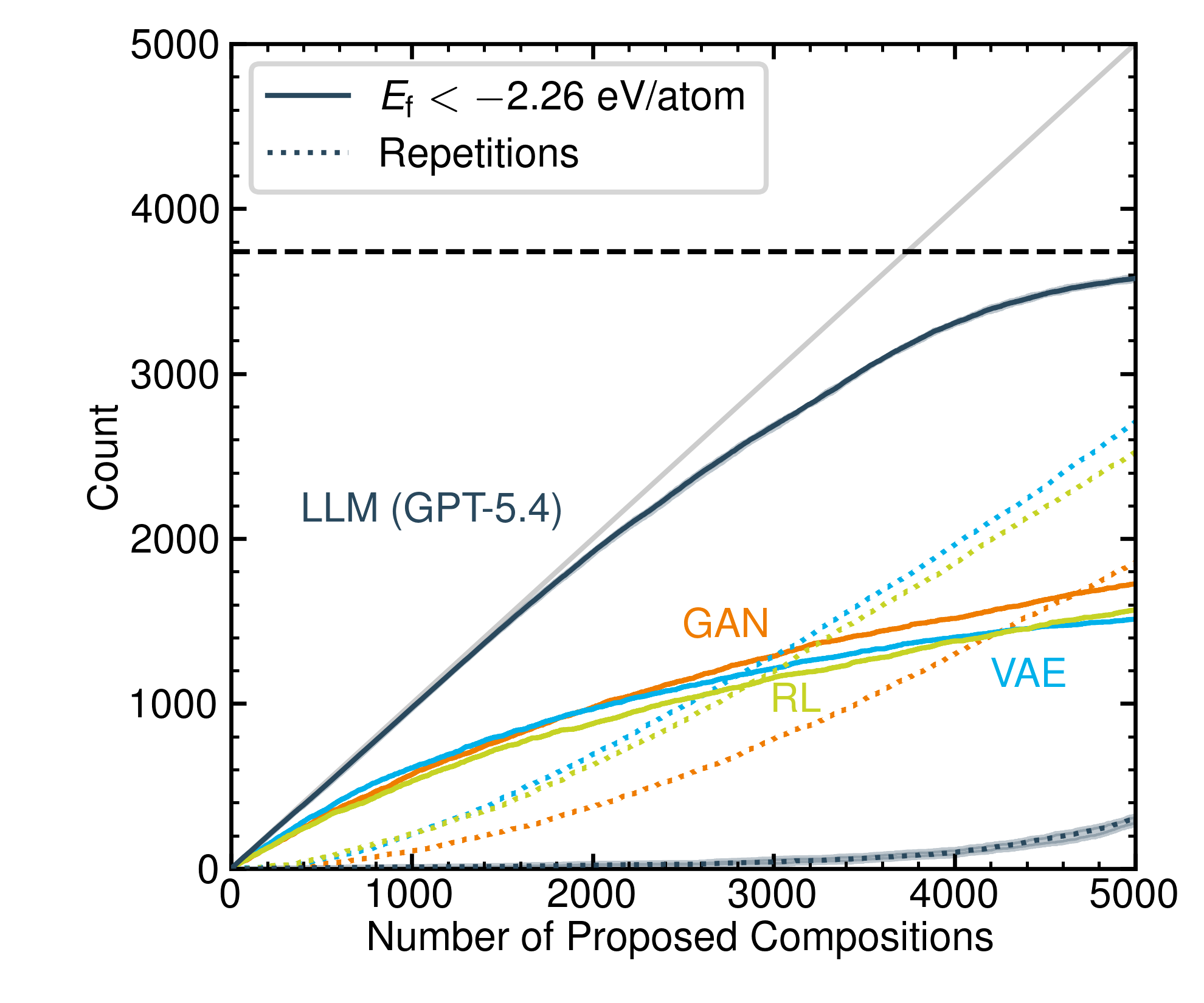}
    \caption{Model performance as a function of the number of generated compositions. Shown are the number of generated compositions with a formation energy below \SI{-2.26}{\electronvolt}/atom (solid lines) and the count of repeated compositions (dotted lines) for the LLM (GPT-5.4, dark blue), the GAN (orange), the VAE (light blue) and the RL model (green) reproduced from Ref.~\cite{turk2022assessing}. The black dashed line shows the limit of 3,740 compositions with $E_\mathrm{f}<$\SI{-2.26}{\electronvolt}/atom. The identity line is displayed in gray. In case of the LLM, the lines denote the average over three runs, while the shaded areas indicate the divergence between best and worst run. }
    \label{fig:performance}
\end{figure}

Figure~\ref{fig:performance} compares the performance of OpenAI's GPT-5.4 model\autocite{singh2025openai} against the three generative models from Türk et al.\autocite{turk2022assessing} with key performance metrics summarized in Table~\ref{tbl:performance}.
Details on the model selection can be found in Section~S2 of the Supporting Information.
The LLM outperforms all three models.
Within 5,000 proposed compositions it identifies on average 3,577 compositions below the target threshold (96\%), with the best run identifying 3,590 compositions.
This performance is particularly remarkable, given that the LLM has received no task-specific training and relies solely on in-context learning from iterative feedback. 
Further, we find that the performance of the LLM is not limited to the most stable compositions, and that other formation energy ranges can be targeted reliably,
too (see Section~S3 of the Supporting Information).

\begin{table}[h]
\small
  \caption{Performance comparison of different generative models in proposing low-formation-energy Elpasolite compositions. The columns report the first occurrence of a repeated composition, the number of Elpasolites with $E_\mathrm{f}<$\SI{-2.26}{\electronvolt}/atom discovered after 5000 samples, and the maximum number identified across all generated compositions ($^*$ 5000 for the LLM ). For the LLM, the average over 3 runs is shown.}
  \label{tbl:performance}
  \centering
  \begin{tabular*}{0.5\textwidth}{@{\extracolsep{\fill}}lllll}
    \hline
    Model & 1. repetition & \# Elp. at 5000 & Max Elp. found  \\
    \hline
    GAN & 35 & 1726 & 3550\\
    VAE & 64 & 1511 & 2977\\
    RL & 91 & 1569 & 2831\\
    LLM & \textbf{297} &\textbf{3577}& \textbf{3577}$^*$\\
    \hline
  \end{tabular*}
\end{table}

A significant advantage of our LLM workflow is the iterative feedback.
At each call, the LLM is provided with a list of all previously proposed compositions.  
In contrast, the other generative models statistically draw samples leading to an early onset of repeated (previously proposed) compositions, already after 35 to 91 generation attempts.
This lack of ''composition memory'' leads to repetitions increasingly dominating the composition generation process after 1,511 (VAE) to 4,677 (GAN) proposed compositions.
While the LLM also repeats compositions, with the first repetition occurring between 159 to 548 proposed compositions in this case, the total number of repeated proposals remains a fraction of the successfully identified low-energy compositions.
The iterative feedback moreover allows the model to reason over the growing history of proposals and their formation energies, continuously refining the generation strategy.
This dual benefit of fewer wasted proposals and increasingly informed suggestions is a key driver of the LLM’s efficiency advantage over the other generative models. These factors are explored in more detail below, where we analyze the key components underlying the LLM’s performance.

We note that, as with any machine-learned reference, the predicted formation energies carry inherent uncertainties that are worth keeping in mind when interpreting the results.
In particular, \ce{ABC2D6}/\ce{BAC2D6} pairs exhibit deviating energy predictions, with 688 pairs falling both above and below the \SI{-2.26}{\electronvolt}/atom threshold -- possibly a consequence of the AB-permuted structures being underrepresented in the original training set and the representation's lack of invariance with respect to such permutations (see Section~S4 of the Supporting Information).
Moreover, 262 compositions below \SI{-2.26}{\electronvolt}/atom contain noble gases.
While it may be conceivable that noble-gas atoms occupy e.g. the C-site in the cages formed by the other sites, it is questionable whether such a material would exist beyond the regression model.
This at first unintuitive fact is reflected in the order in which the LLM proposes certain elements, with the noble gases only being introduced after around 3,100 generated compositions, as further illustrated in Section~S4 of the Supporting Information.
The LLM’s apparent deficiencies relative to the nominal 3,740 targets may therefore partly be attributed to inaccuracies in the reference data, and the true performance could potentially be even better.

\subsection{Model size}

It is well established that model size plays a critical role for performance. 
Typically, larger models perform better across a wide range of tasks but smaller models are usually faster and cost a fraction of their larger counterparts.
We therefore systematically assess the influence of model size on the generation performance.
To this end, we evaluate two smaller models, GPT-5.4-mini and GPT-5.4-nano, and compare their effectiveness when tasked with generating 5,000 candidate compositions.

\begin{figure}
    \centering
    \includegraphics[width=0.6\linewidth]{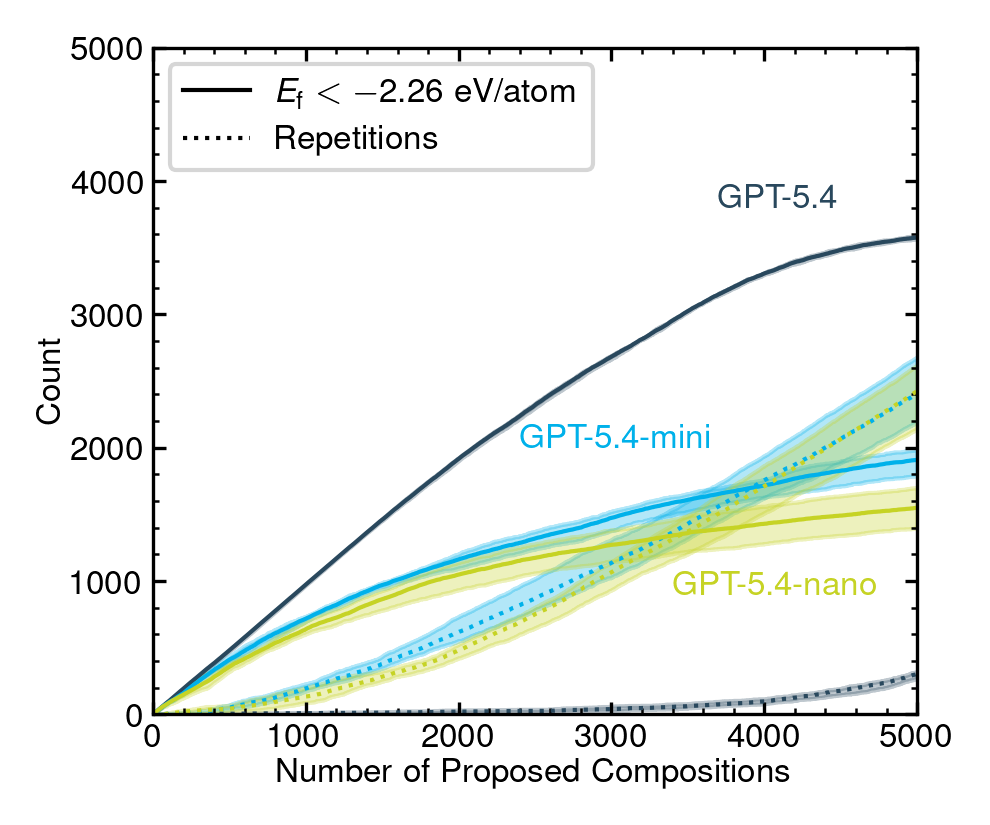}
    \caption{Composition generation performance as a function of the number of generated compositions, comparing different model sizes of the LLM. Shown are the number of generated compositions with a formation energy below \SI{-2.26}{\electronvolt}/atom (solid lines) and the count of repeated compositions (dotted lines) for GPT-5.4 (dark blue), GPT-5.4-mini (light blue), and GPT-5.4-nano (green). The lines denote the average over three runs, while the shaded areas indicate the divergence between best and worst run.}
    \label{fig:model_size}
\end{figure}

As expected, and illustrated in Figure~\ref{fig:model_size}, the smaller models exhibit markedly inferior performance. 
Although the gap between the mini and nano models is modest, both are significantly outperformed by the larger model. 

The most pronounced deficiency of the smaller models is the higher occurrence of repetitions.
For GPT-5.4-mini, the number of repeated proposals exceeds the number of newly identified low-energy compositions after approximately 3,300 to 4,500 generated samples. 
A similar trend is observed for GPT-5.4-nano, where this occurs even earlier, between around 3,000 and 4,100 samples. 
This behavior indicates that smaller models struggle to effectively utilize or retain contextual information over long generation sequences.
In practical terms, they appear to “forget” previously generated compositions, leading to redundant output and inefficient exploration of the search space.

This limitation is especially detrimental in large compositional spaces, where maintaining diversity and avoiding repetition is crucial for discovering novel low-energy candidates.
Consequently, these findings suggest that larger (general-purpose) models are currently necessary to reliably explore large compositional spaces, as they are better equipped to handle long-context dependencies and maintain effective search behavior.

However, this conclusion should be interpreted in light of the intended application.
If only a limited number of compositions is required — for example, on the order of 25 candidates — the performance of smaller models may well suffice.
As demonstrated in Section~S5 in the Supporting Information, under such constrained conditions the drawbacks associated with repetition and context degradation are less pronounced, making smaller models viable and cost-effective alternatives.

\subsection{Reasoning}

\begin{figure}[h!]
    \centering
    \includegraphics[width=0.6\linewidth]{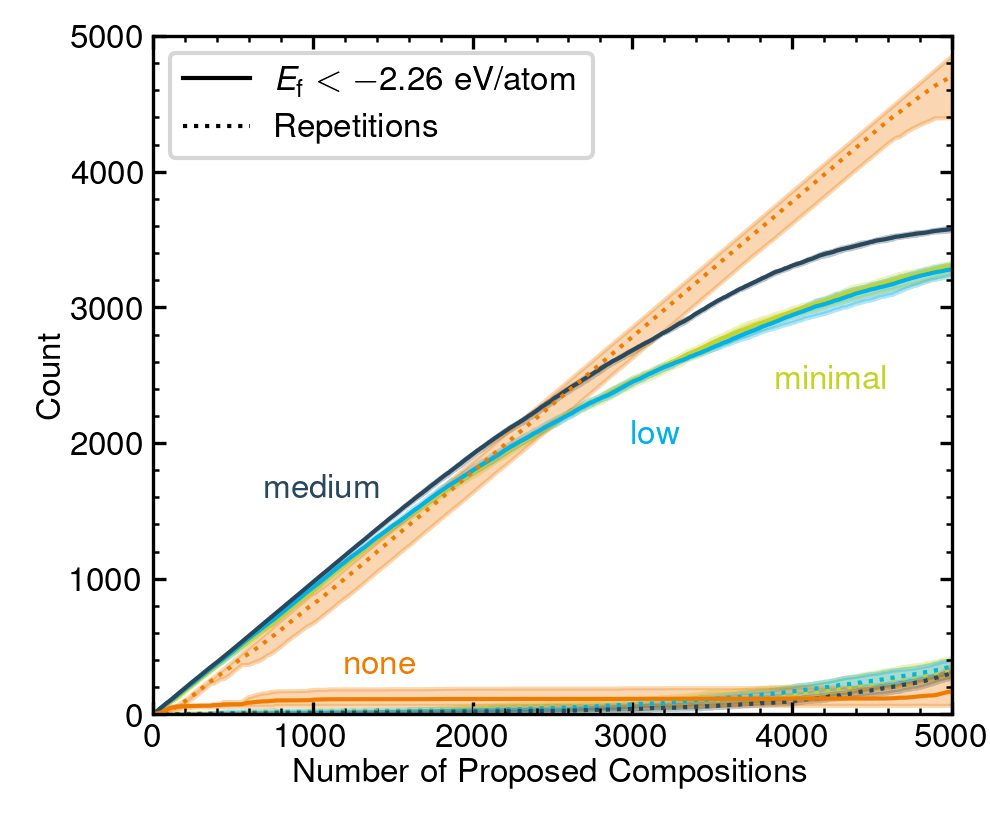}
    \caption{Composition generation performance as a function of the number of generated compositions, comparing different reasoning effort levels. Shown are the number of generated compositions with a formation energy below \SI{-2.26}{\electronvolt}/atom (solid lines) and the count of repeated compositions (dotted lines) for medium (dark blue), low (light blue), minimal (green), and no (orange) reasoning effort. The lines denote the average over three runs, while the shaded areas indicate the divergence between best and worst run.}
    \label{fig:reasoning}
\end{figure}

Prompting LLMs to reason through intermediate steps has been shown to substantially improve their performance.\autocite{wei2022chain}
In earlier composition generation frameworks, agentic tools combining LLMs with structural data from the Materials Project and stability prediction models, the model's reasoning was guided through explicit prompting instructions.\autocite{jia2024llmatdesign}
In contrast, many recent API-based models automate this and allow flexible control of reasoning depth via parameters such as \texttt{reasoning\_effort}.
In Figure~\ref{fig:reasoning}, we quantify the effect of varying the allocated number of reasoning tokens on model performance.
At medium effort, reasoning tokens constitute 7\% of all tokens consumed per run.
This fraction drops to approximately 1.5\% for low and minimal effort, and to zero when reasoning is disabled entirely.
Since output tokens are meaningfully more costly than input tokens, this modest fraction however has a pronounced effect on the total expense.

Across a set of 5,000 generated compositions, the model configured with medium reasoning effort identifies on average 96\% of Elpasolite structures with formation energies below $E_\mathrm{f} < \SI{-2.26}{\electronvolt\per atom}$.
In comparison, low and minimal reasoning effort configurations both achieve 88\% coverage, indicating a slight dependence of output quality on reasoning depth.
Notably, even minimal reasoning yields competitive performance, achieving a substantial fraction of high-quality candidates at only 65\% of the cost associated with medium reasoning, as displayed in detail in Section~S6 of the Supporting Information.
This suggests that reduced reasoning effort may offer an efficient alternative with only moderate performance degradation depending on the application and resource constraints.
At the same time, disabling reasoning entirely leads to a marked drop in performance, underscoring that a certain degree of reasoning is essential for this task.
For the present study we adopt a medium reasoning setting.
Higher reasoning levels (e.g.\ \texttt{high} or \texttt{xhigh}) are expected to yield further modest improvements, albeit at higher cost.

\subsection{Starting composition}

Prompt design plays a crucial role in an LLM’s effectiveness.
A common technique to improve LLM performance is using one-shot or few-shot prompting, in which an LLM is given one or more examples of the target task before being asked to complete a new, similar instance.
Such prompting strategies have been shown to improve model performance.\autocite{brown2020language}

We investigate three different one-shot prompting modes, each employing a different starting composition: the prototypical Elpasolite mineral \ce{AlNaK2F6}, the anonymous formula \ce{ABC2D6}, and the composition with the highest predicted formation energy in the dataset, \ce{ArKrN2C6}.
Importantly, we modified the prompt, so that no explicit information about the underlying crystal structure --- i.e., that the compositions correspond to the Elpasolite structure type --- was provided in any of these prompts, allowing us to isolate the effect of the starting composition on generation performance (see Section~S7 in the Supporting Information). 

\begin{figure}[h!]
    \centering
    \includegraphics[width=0.6\linewidth]{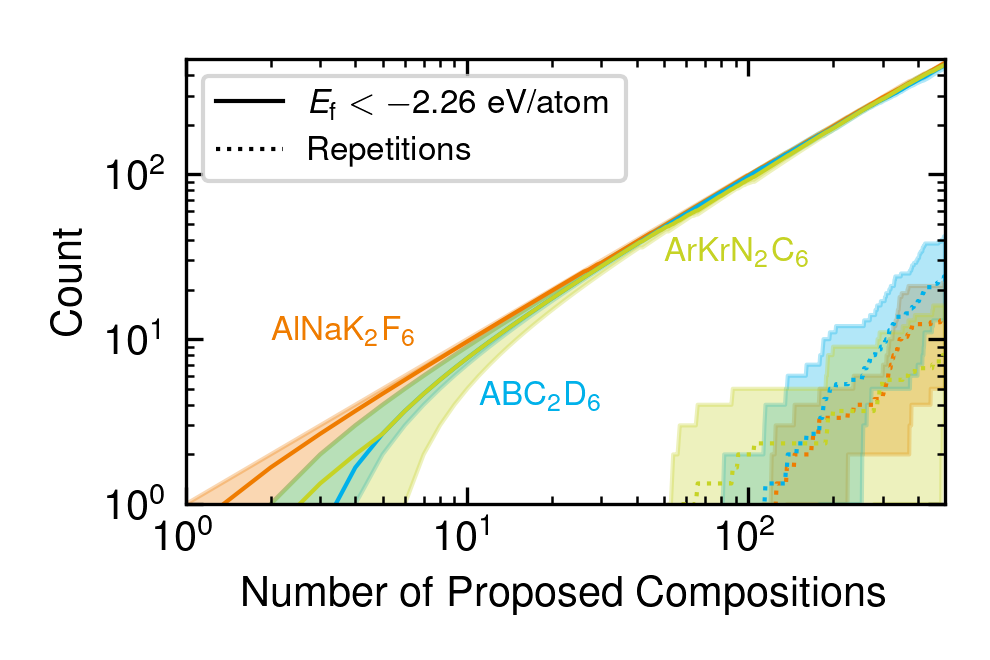}
    \caption{Composition generation performance as a function of the number of generated compositions, comparing different starting compositions and without mentioning the term Elpasolite in the prompt. Shown are the number of generated compositions with a formation energy below \SI{-2.26}{\electronvolt}/atom (solid lines) and the count of repeated compositions (dotted lines) for the initial compositions \ce{AlNaK2F6} (orange), \ce{ABC2D6} (light blue), and \ce{ArKrN2C6} (green). The lines denote the average over three runs, while the shaded areas indicate the divergence between best and worst run.}
    \label{fig:one-shot}
\end{figure}

As shown in Figure~\ref{fig:one-shot}, the choice of starting composition has a moderate impact on early-stage performance.
Initializing the model with the realistic prototype \ce{AlNaK2F6} leads to a noticeable improvement in the quality of generated compositions within the first few proposals, likely because the example immediately activates relevant chemical knowledge encoded during pre-training.
Interestingly, even when initialized with the anonymous composition \ce{ABC2D6}, the model demonstrates a form of emergent ''chemical intuition'': despite the absence of any explicit structural information, the LLM is capable of proposing compositions with low formation energies.

A key factor underlying this behavior is the LLM’s ability to assign chemically plausible elements to the corresponding lattice sites.
As a case in point, the LLM may follow common chemical conventions, such as ordering elements by electronegativity, thereby placing highly electronegative elements at the D site in the Elpasolite structure. 
As illustrated in Section~S7 of the Supporting Information, the model consistently selects fluorine as the anion on the D site in all three cases, correctly identifying it as the anion in our search space that yields 99.7\% of compositions below the $\SI{-2.26}{\electronvolt\per atom}$ threshold.
Furthermore, the model appears to recognize that varying especially the elements on sites A and B is a suitable way to explore the low formation energy region of the Elpasolite space.

In Section~S8 of the Supporting Information, we further demonstrate that this behavior persists even when the composition search is fully abstracted.
Using an anonymous formula representation, the model continues to generate chemically meaningful candidates by inferring the site dependencies and the physical scale of the formation energies from the iterative feedback alone.
This behavior demonstrates the reliability and transferability of this approach especially when applied to completely new systems, as the LLM can capture patterns that are not yet present in the literature.

\subsection{Iterative feedback}

\begin{figure}[h!]
    \centering
    \includegraphics[width=0.6\linewidth]{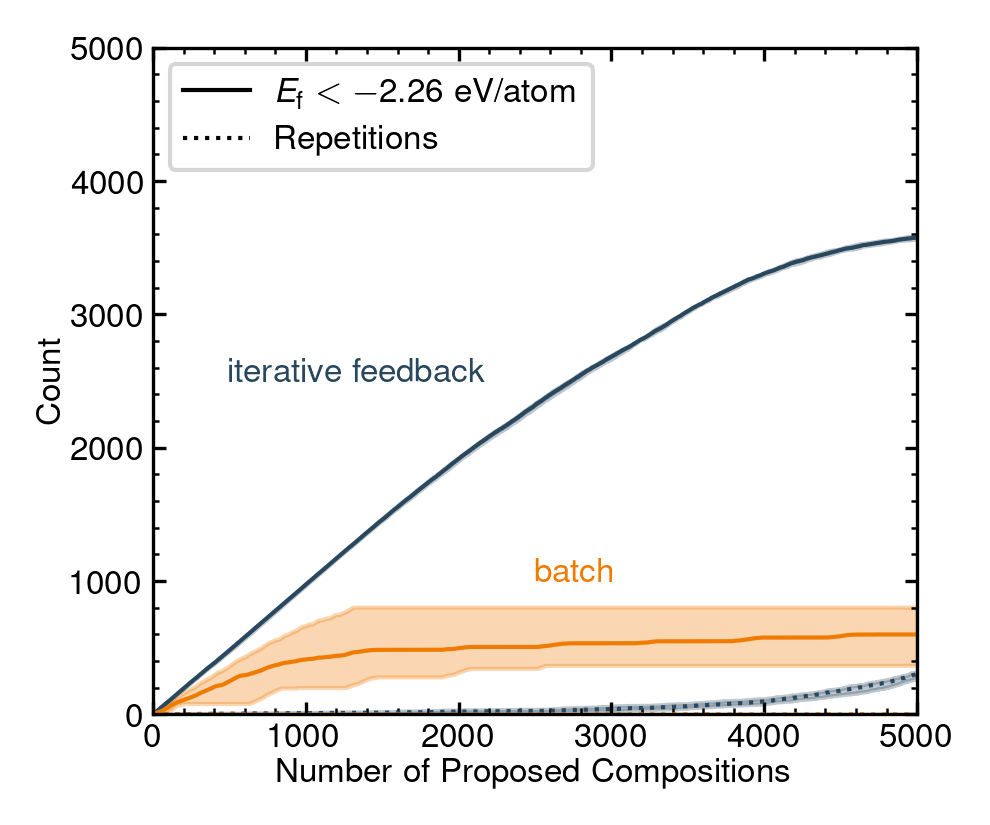}
    \caption{Composition generation performance as a function of the number of generated compositions, comparing ''iterative''- (dark blue) and ''batch''- (orange) prompting. Shown are the number of generated compositions with a formation energy below \SI{-2.26}{\electronvolt}/atom (solid lines) and the count of repeated compositions (dotted lines). The lines denote the average over three runs, while the shaded areas indicate the divergence between best and worst run.}
    \label{fig:batch}
\end{figure}

To further probe the ''chemical intuition'' exhibited by the model, we consider a setting in which iterative feedback is withheld entirely: instead of receiving the outcome of each proposal before making the next, the LLM is prompted to generate a large batch of 5,000 candidate compositions in a single pass.
Figure~\ref{fig:batch} shows that this batch-generation mode leads to a substantial drop in performance compared to the iterative approach and highlights the importance of feedback in guiding the model toward chemically meaningful regions of the compositional space.

\begin{figure}
    \centering
    \includegraphics[width=0.5\linewidth]{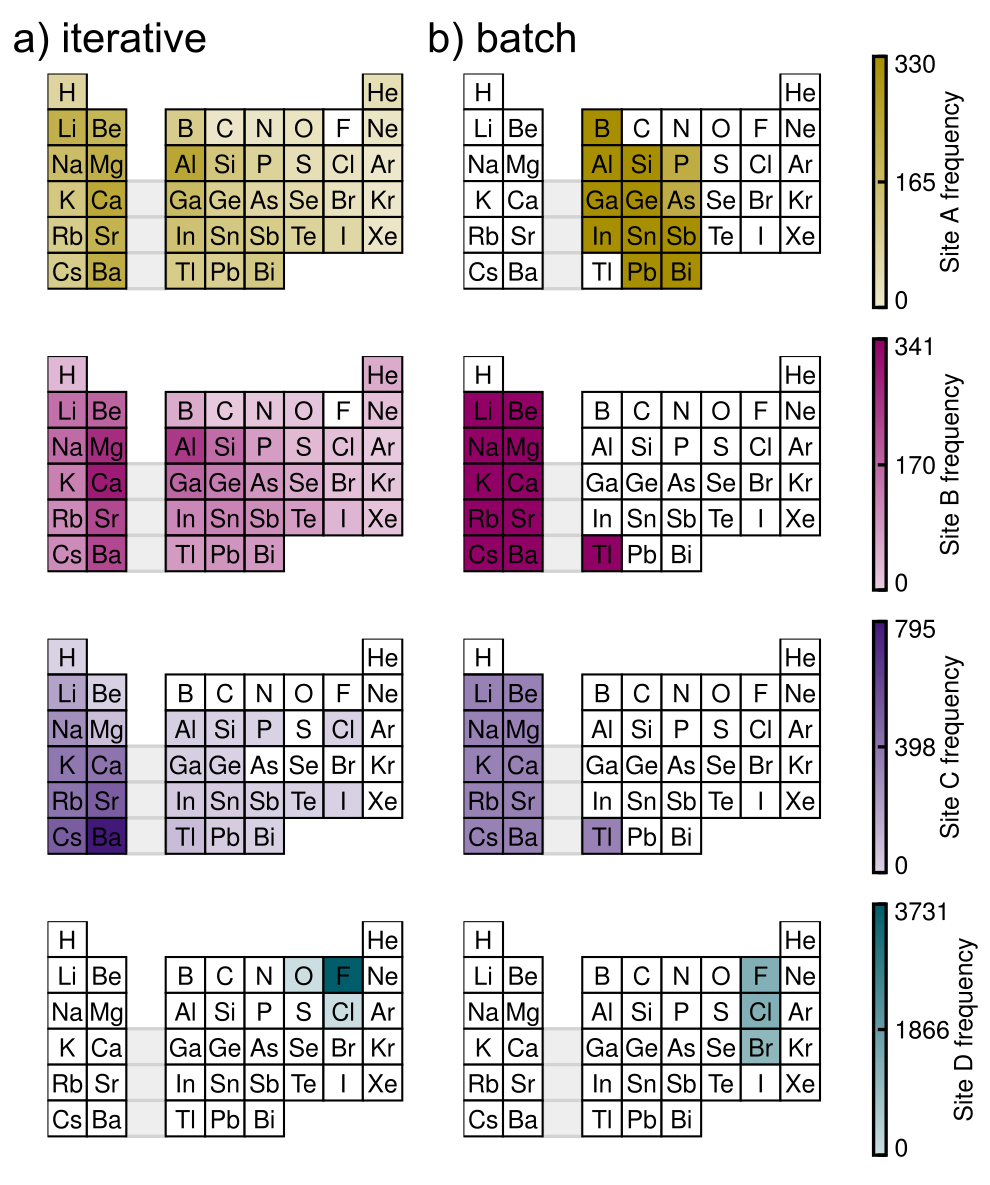}
    \caption{Element distribution on each site for the first 3740 generated Elpasolite compositions, for both the best performing a) iterative and b) batch run.}
    \label{fig:pse_2}
\end{figure}

Without such feedback, the model’s proposals become significantly less targeted, as is illustrated in Figure~\ref{fig:pse_2}.
In this batch mode, the model systematically substitutes the prototype composition \ce{AlNaK2F6} with elements occupying similar positions in the periodic table.
While this still reflects the model’s inherent ''chemical intuition'', it illustrates how effectively the model applies in-context learning in the iterative mode.
For sites A, B, and C, this results in a broader exploration of the periodic table compared to the batch mode. 
Further, the model quickly learns through iterative feedback that fluorine promises the most success as the D site occupant. While other halogens might initially also appeal as D site anions, they ultimately lead to higher formation energies than fluorine. 
The choice of noble gases further illustrates how this iterative approach lets the model explore at first unintuitive options on-the-fly, as further illustrated in Section~S9 of the Supporting Information. While this might be an error of the underlying dataset as previously discussed, it showcases how LLMs can navigate such unconventional compositions in a targeted manner.

Importantly, the observed performance gap also provides evidence that the model does not merely recall any Elpasolite data from its pre-training data.
If that were the case, one would expect similarly strong performance even in the batch-generation mode.
Finally, we note that the iterative setup with feedback effectively introduces an element of active learning, constituting an advantage over the baseline generative models that operate in a purely one-shot manner.
This asymmetry in the generation architecture should be taken into account when comparing performance across methods.
However, the simplicity by which such active learning-like behavior can be achieved in LLMs is a strong benefit of this approach.

\FloatBarrier
\subsection{Iterative batch mode}

Lastly, we explore a hybrid strategy that combines the advantages of iterative feedback with the cost efficiency of batch generation.
In this mode, the LLM is prompted to propose a small batch of compositions at once. The formation energies for the entire batch are then evaluated and returned collectively as feedback, after which the model proposes the next batch of the same size.
This procedure enables iterative refinement with substantially fewer feedback cycles than purely sequential generation.

\begin{figure}
    \centering
    \includegraphics[width=0.6\linewidth]{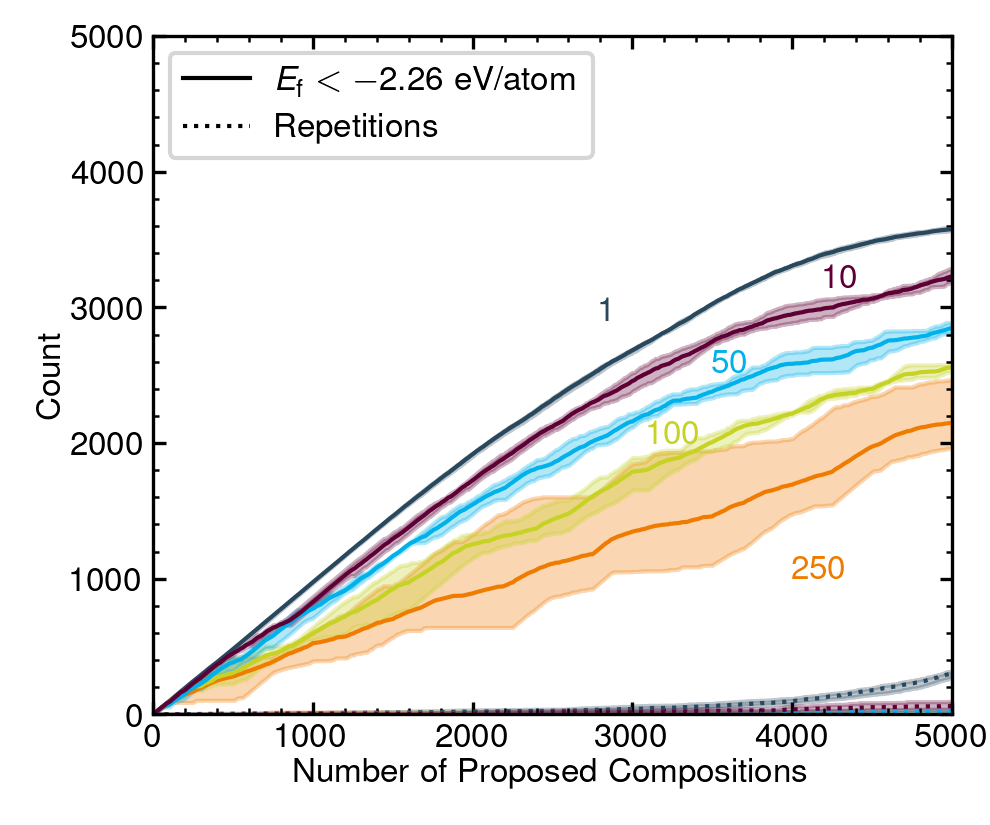}
    \caption{Composition generation performance as a function of the number of generated compositions, comparing ''iterative''- (dark blue) and ''batch''-prompting with increasing sizes of generated batches. Shown are the number of generated compositions with a formation energy below \SI{-2.26}{\electronvolt}/atom (solid lines), above \SI{-2.26}{\electronvolt}/atom (dashed lines) and the count of repeated compositions (dotted lines). The lines denote the average over three runs, while the shaded areas indicate the divergence between best and worst run.}
    \label{fig:iter_batch}
\end{figure}

Figure~\ref{fig:iter_batch} shows performance for batch sizes of 10, 50, 100, and 250.
As expected, increasing the batch size leads to a gradual degradation in performance, since the model receives feedback less frequently and thus has fewer opportunities to adapt its proposals.
Nevertheless, batch sizes in the range of 10--50 retain a substantial fraction of the performance of the full iterative mode while significantly reducing the number of LLM calls.
The runtime and cost for each batch size are shown in Section~S10, however we expect this to become less critical as models continue to improve in efficiency and inference costs decrease.
By carefully selecting the batch size, one can therefore achieve a favorable trade-off between sample efficiency (hit rate) and computational efficiency (number of LLM calls and property evaluations).
This approach is particularly advantageous when property evaluation is fast and inexpensive, enabling efficient exploration of large compositional spaces with reduced overhead from iterative model querying. 
More involved strategies that sub-select the most important compositions to evaluate in each batch could further allow to reduce the sampling cost in case of a more expensive property evaluation.

\section{Discussion}

We have demonstrated that a general-purpose LLM without task-specific fine-tuning can serve as an effective constrained composition generator for inorganic crystals in large design spaces.
Using the Elpasolite dataset as a pre-tabulated benchmark, our iterative feedback framework achieves performance surpassing purpose-trained generative models (GAN, VAE, RL), identifying on average 3,577 out of 3,740 (96\%) low-formation-energy compositions within just 5,000 proposals.
This result is particularly striking given that the purpose-trained models require substantial training overhead and still saturate at lower discovery rates.

A central finding is that in-context learning from iterative feedback drives the model's strong performance.
When feedback is withheld and the LLM generates compositions in a single batch, performance degrades substantially, which also rules out the possibility that the model simply recalls the target set from its pre-training data.
Instead, the model appears to leverage its broad chemical knowledge to progressively refine its compositional strategy in response to the growing history of proposals and their formation energies.
This is further evidenced by the model's emergent chemical intuition: even without being told the material class, it reliably identifies fluorine as the occupant of the D-site and selects various chemically appropriate cations for each crystallographic position.
We further find that the allocated reasoning effort has a meaningful impact on performance.
While even minimal reasoning yields competitive results, disabling it entirely leads to a marked drop in performance, underscoring that some degree of reasoning is indispensable.

A key practical advantage of LLM-based generation is its flexibility and the ease with which constraints can be imposed.
Maintaining uniqueness effectively, meaning that repeated compositions should explicitly be avoided by the LLM, is another advantage over other deep generative models.
As a case in point, the three generative models investigated by T\"urk et al. predominantly generate repetitions after roughly 4,000 generated compositions, while the LLM mostly suggests unique compositions until the end.  
Similarly, physical and chemical constraints, such as charge neutrality, electronegativity balance, site occupancy rules, or elemental exclusions, could be enforced directly through the prompt without needing to modify the model.
Switching to a new material class or target property requires only an adapted prompt, and the approach benefits automatically from future LLM improvements.
This stands in contrast to specialized generative models, which must be retrained from scratch when the task changes.

Some limitations remain and must be considered.
The cost scales quadratically with the number of iterations as the full proposal history is input at each step, which motivates the iterative batch strategy as a practical mitigation.
The context window imposes a hard upper bound on the history that can be retained, and performance may degrade for material classes or element combinations that are underrepresented during the model's pre-training.
Moreover, the framework also benefits from a fast property calculator (in our case a pre-tabulated dataset).
Direct integration with expensive first-principles calculations would, thus, require a surrogate model, for instance one of the increasingly popular, universally applicable ML potentials.\autocite{batatia2025foundation}
Finally, there is no guarantee of systematic coverage: the model may over-exploit chemically familiar regions, leaving other viable areas of the composition space unexplored.

Overall, our results establish general-purpose LLMs as a useful and accessible alternative to dedicated generative models for constrained materials composition search at scales ranging from hundreds to thousands of candidate compositions, and underscore their potential as a flexible component in agent-driven materials discovery pipelines.

\section*{Acknowledgements}

We thank David Greten for the technical discussions and feedback on the LLM workflow presented herein.
We thank Christian Kunkel for fruitful discussion on the GAN, VAE and RL models.
We furthermore thank Christoph Scheurer for initial discussions on agentic pipelines as well as organizational support regarding OpenRouter accounting.
This work uses computational resources of the Max Planck Computing and Data Facility (MPCDF).

\section*{Author Contributions}
Author contributions below follow the Contributor Roles Taxonomy (CRediT):

\noindent \textbf{Hedda Oschinski}: Conceptualization, Software, Formal analysis, Investigation, Data Curation, Writing - Original Draft, Writing - Review \& Editing, Visualization; \textbf{Maximilian L. Ach}: Conceptualization, Software, Formal analysis, Investigation, Data Curation, Writing - Original Draft, Writing - Review \& Editing; \textbf{Konstantin S. Jakob}: Conceptualization, Writing - Review \& Editing; \textbf{Christian Carbogno}: Conceptualization, Writing - Review \& Editing, Supervision; \textbf{Karsten Reuter}: Conceptualization, Resources, Writing - Review \& Editing, Supervision

\section*{Conflicts of interest}

The authors declare no conflict of interest.

\section*{Data Availability}

The code for the workflows, all data, and all results (including LLM prompt outputs) required to reproduce the findings will be made available upon acceptance.

%%%END OF MAIN TEXT%%%

%%% Bibliography
\printbibliography

\end{document}